# Proposal for a quantity based data model in the Virtual Observatory

Brian Thomas[a,b] and Edward Shaya[a,c,d]
Code 630.1
Goddard Space Flight Center/NASA
Greenbelt, MD 20012

## Abstract

We propose the beginnings of a data model for the Virtual Observatory (VO) built up from simple ``quantity'' objects. In this paper we present how an object-oriented, domain (or namespace)-scoped simple quantity may be used to describe astronomical data. Our model is designed around the requirements that it be searchable and serve as a transport mechanism for all types of VO data and meta-data. In this paper we describe this model in terms of an OWL ontology and UML diagrams. An XML schema is available online.

## 1. Introduction: an object- and domain-oriented approach

The VO community is currently in the process of attempting to formulate a data model which might be shared across all data repositories and used to facilitate the query, exchange and fusion of astronomical data. This fundamental requirement must also be coupled with the need for this data model to be able to replicate the structure and content of the data at any participating VO data repository. There have been a number of prior initial attempts to formulate such a data model, with perhaps the most mature and insightful being that put forth by McDowell etal, [1,2]. In the McDowell model a number of meta-data and data challenges are outlined and an initial model proposed (see figure 7, 8 of [1]). One of the major advancements of this model is that meta-data and data are to be encapsulated in an object-oriented fashion, similar to the manner in which XDF (``eXtensible Data Format'', [3]) handles its information. The object-oriented approach allows for extension of existing concepts into new concepts which inherit the attributes of their parent concepts. With such methodology, it is possible to reuse important concepts and perhaps allow for a mechanism whereby a computer may decompose advanced child concepts into more digestible parent concept parts. Unfortunately, the McDowell model appears to join all of the meta-data concepts into a single domain, that of the data model itself. As Plante [4] has argued this kind of approach will make for an increasing problem both for initial development and long-term maintenance of the VO data model meta-data. If we however sever the connection of the data model to the astronomical meta-data we will allow for much greater progress on the problem of developing the data model and its maintenance.

[a]also Raytheon Technical Services Company, 4500 Forbes Blvd., Lanham MD
[b]email: brian.thomas@gsfc.nasa.gov
[c]email: shaya@mail630.gsfc.nasa.gov
[d]also Department of Astronomy, University of Maryland, College Park, 20740

In short summary then, what is needed by the VO community is a new data model which contains the ability to be extended in an object-oriented fashion, has domains which may allow for separate communities within the VO to apply their expertise without stepping on other areas of work, and has the ability to be used as an exchange format across the Internet.

Recently Plante [5] has argued for a quantity-based data model, where a quantity is used to model all of the data interactions. We have taken this idea in hand to develop our proposal put forth in this paper based on our own work with XDF and guided by the work of many others in the VO (refs above plus: VOTable, [6,7]; VO data model mailing list [8]; UCD [9], SIMD [10]) and larger astronomy communities (FITS [11]). In this paper  we propose the beginnings of such a data model and describe how adopting object-, domain-, and quantity-oriented approach may serve to promote a sharable, searchable standard amongst the various data repositories of the VO.

*High-level requirements and design philosophy*

To serve as a basis for our modeling we will want to address some specific, important requirements that we feel any version VO data model should. The highest level requirement, that the data model be capable to facilitate the query, exchange and fusion of astronomical data may be broken out into the following more specific requirements:

1   The data model should be capable to hold as many types of data as possible ranging from numbers to algorithms which generate numbers. The datum held by the model can be either scalar , vector  numerical values or structures of other objects (quantities).

2   The model may support complex data arrangements, including, but not limited to single values, arrays of values, images, tables with an admixture of fields (each with different units, data format and accuracy) and n-dimensional data cubes.

2   The data model may be incorporated into XML so that it may serve as a building block for creation of a searchable interface document of an archive and as a web-ready transport format.

Plus, as our own design principle, the data model should be as spare as possible but its few components immanently reusable and extensible. It should contain NO meta-data which are not absolutely  needed (such as meta-data from specialized sub-fields of astronomy).

**2. Theoretical considerations**

## 2.1 Definition of quantities

We start by considering a "Quantity" exists pairs a "concept" with its "value". The concept may be any meaningful term or idea which the VO community wishes to use (for example "X-ray star", "visual flux", "CCD camera type", "index", etc) and the value indicates the amount or degree of the concept. The value may be a number, a string (such as "high", "big", "large") or a symbol.

To serve as a general framework for manipulating concepts and their values we create an entity or object that serves as the package for all of these tuples. This entity, the "quantity" is inherited (passes on all of its properties) to all concepts (Figure 1) and we thus tie the meaning to the class of the object.

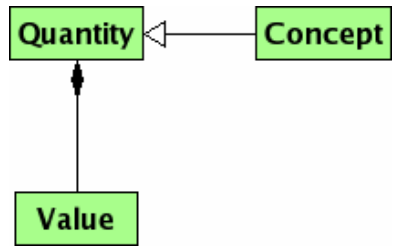

**Figure 1.** *A simple UML diagram of the relationship between a concept, the quantity and its value.*

 The value is itself a one-dimensional array, which may have one or more "datum", e.g. the value V may be defined:

$$V = \{ d_1, d_2, d_3, .. d_n \} \qquad\qquad (1)$$

where n is the number of datum in V and each $d_i$ are the individual datum that may be either scalars, vectors or other quantities. In the last case the child quantity serves as data but we may also use child quantities to specify meta-data. To do so we insert the child quantity directly within the parent quantity with the designated relationship "metaData"(figure 2).

As the datum in V must describe scientific information (at least most of the time) each datum must have associated with it some description of its accuracy (errors) and scientific units.  Machine understandability requires that each datum should also be described by some type of a data format. In our model, we consider that all datum in V for a given Q are "homogeneous" meaning *that every datum of a given Q has the same units and data format*. Accuracy of scientific data can, and often does, vary on a datum by datum basis. Thus, we infer the existence of an array of accuracy values  which is of the same size as the array of V to which it refers.

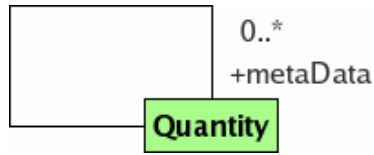

***Figure 2.*** *Quantities may include other quantities as meta-data.*

## 2.2 Mapping quantities into higher dimensional structures – scalar case

Let us first consider the simplest case when the datum of a quantity are scalar and one has a 1-dimensional array of them. To create a higher dimensional quantity consider that the datum of the quantity Q (arranged along index $k$) may be mapped from its linear array V into a higher dimension representation e.g.

$$Q(k) \rightarrow Q(i, j) \qquad\qquad (2)$$

For example, let Q have 16 datum and be a function of i and j which are orthogonal indices. Then for any [i, j] location there is then a value $Q_{ij}$. The plane and mapped locations looks like that in figure 3.

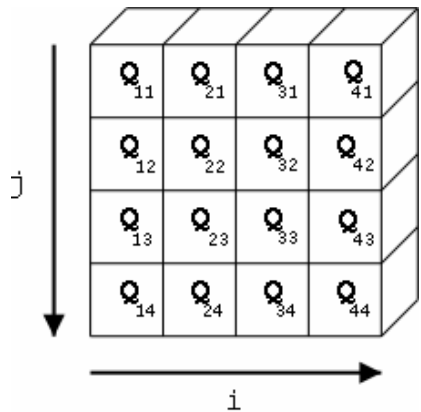

***Figure 3.*** *Virtual locations of $Q_{ij}$.*

Now it is clear that the indices $i$ and $j$ describe not only "arguments"of (or, alternatively, "dependencies") but also dimensions of Q, with the values on each dimension numbering 1 to n where n is the size of that dimension. As Q is a function of i and j, we call i and j "arguments"of Q.

Now we may associate each of our datum in V to this virtual plane by adopting some mapping. One example, where the first dimension specified in the function is the

"faster"one as shown in figure 4.

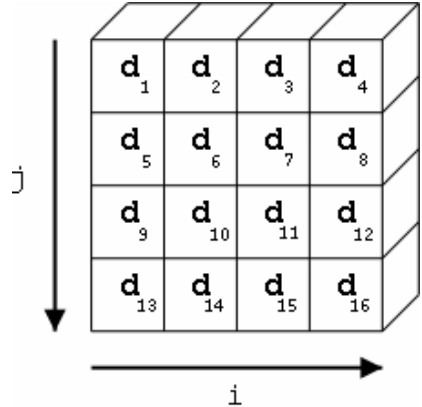

**Figure 4.** *Suggested mapping between datum and virtual plane locations.*

It is easy to see that we need not stop at 2 dimensions. We may map the linear array of datum in V to any set of the same or higher dimensional "virtual"locations. We thus consider that for any set of arguments, the quantity Q may be mapped to a "data cube"(a hyper-plane) of dimension equal to the number of arguments it has (e.g. the quantity "Q(a,b,c,d)"is a 4-dimensional "cube"of datum). To keep things consistent between the virtual data cube and V we require that the number of datum in Q be equal to the multiple of the size of its dependent dimensions. Thus, if Q has no arguments it is of 0 dimensional extent, and may hold only a single datum in its linear array V.

Now taking up the plane in figures 3 and 4 consider that the axes of *i* and *j* are nothing more than quantities themselves, albeit simple quantities which take 'integer' scalar datum and have the restrictions that the values are limited to being whole numbers. We would then define these simple quantities as functions of the indices *i* and *j* using the notation:

$$q_i = q1 \ (i)$$

$$q_j = q2 \ (j)$$ 

$$(3)$$

Now we see that the parent quantity Q depends on the quantities $q_i$, $q_j$, or

$$Q = Q \ (q_i, q_j) = Q \ ( \ q1(i) \ , q2(j) \ ) \qquad (4)$$

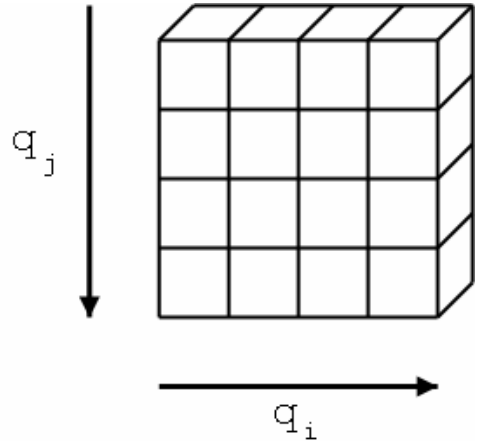

***Figure 5.*** *A data cube where locations are described by quantities  $q_i$, and $q_j$.*

A 2-dimensional data cube would look like that in figure 5 where $q_i$, and $q_j$ are respectively functions of the indices *i* and *j* and each cell in the cube holds a datum of the same units and data format. Now, as each of the quantities  $q_i$, and $q_j$ are functions of a single argument, they are 1-dimensional in extent, and lend their datum as the labels for each of the indices *i* and  *j*. From figure 5 we see that the values of $q_i$ are the scalar set:

$$q_i = \{ 5.1,\ 10.1,\ 34.3 \} \qquad (5)$$

### 2.3 Mapping when the quantity holds vectors

As we mentioned earlier, the datum of a quantity may be vectors. How does that change our picture of the mapping of the datum of the parent quantity? Consider that we might create the vector quantity **L**, which lists a set of sky positions (figure 7) in terms of vectors with components **RA** and **DEC**.

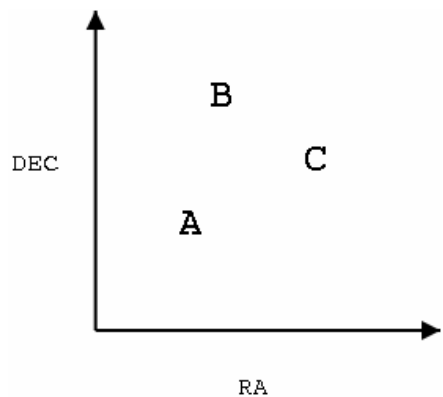

***Figure 7.*** *Example sky position data set.*

Because there are 3 locations we wish to describe, then we may deduce that L must have at least one other quantity as an argument to describe the axis along which the vectors reside. In our case, we choose the quantity "N"(name) which is a scalar quantity that holds a simple string. e.g.

$$L = L\ (N)\quad \text{where } N = \{\ \text{"A", "B", "C"}\}\qquad (6)$$

and the data cube then looks like that shown in figure 8 where we easily may inspect its datum to see that the value of L, at position "A"is a sky vector with component magnitudes "10.2 (ra), 30.1 (dec)".

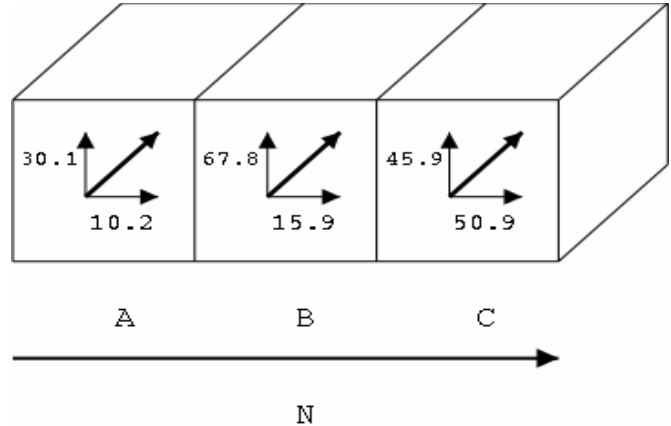

***Figure 8.*** *Data cube of vector data example.*

This example is also interesting as it shows how a multi-variate quantity may be 'indexed' by a single variate quantity. This is useful to shrink dimensionality of a quantity, such that

$$Q\ (\ q_1,\ q_2\ ) \rightarrow Q\ (\ q_3\ )\qquad (7)$$

because $q_3$ is a vector with components $q_1$ and $q_2$.

## 2.4 Mapping values when they are quantities

So far we have described how a multi-dimensional quantity might be created for both scalar and vector data. Now we want to extend the model further to include creation of complex objects such as might occur in tables e.g they have an admixture of various columns each with its own format, accuracy and units. Consider the data shown in Table 1 which comprises 4 columns that are respectively labeled "Name", "$F_{v=2695}$", "Right Ascension", and "Declination"(data taken from [12]).

Notice that the datum of $Q_{set}$ in this case are **not** homogeneous. Each of the quantities N, F, Ra, De controls the data format and units of any particular datum in the set of values for $Q_{set}$.

Table 1. Example tabular data.

| Name | $F_{\nu=2695}$ | Right Ascension | Declination |
|------|----------------|-----------------|-------------|
| 0007+332 | 228.26 | 00 07 49.8 | +33 12 58 |
| 0012+333 | 122.12 | 00 12 09.8 | +33 23 30 |
| 0017+330 | 63.94 | 00 17 21.1 | +33 03 14 |

We consider that each of these columns are essentially independent quantities which are grouped together to form the values of the parent quantity. We thus see the parent quantity, "$Q_{set}$", comprises these quantities, e.g.

$$Q_{set} = \{ \ N, \ F, \ RA, \ De \ \} \qquad (8)$$

where N, F, RA, and De respectively are quantities that represent the columns "Name", "$F_{\nu=2695}$", "Right ascension", and "Declination". Furthermore, we see that the table is laid out in terms of "rows" and we see that all quantities in $Q_{set}$ are mutually dependent on the quantity "row" ("$I_{row}$"). Because this dependence exists, we may also then state that $Q_{set}$ is a function of $I_{row}$ itself and the data cube of $Q_{set}$ is shown in figure 9.

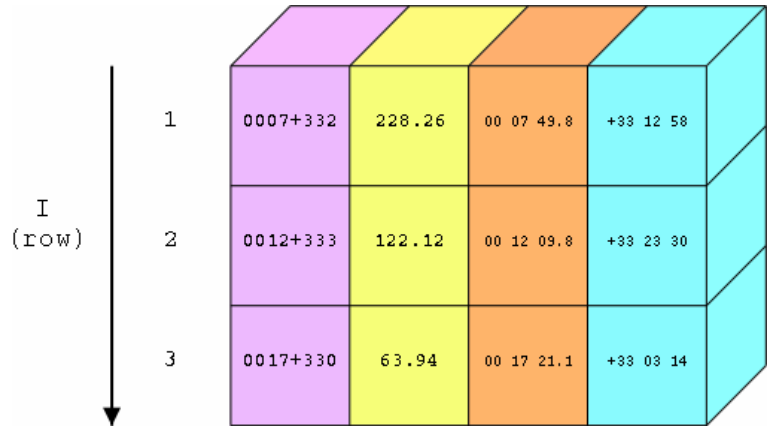

**Figure 9.** *Data cube for table quantity $Q_{set}$ which depends on $I_{row}$. Each colored set of cells represents datum from one of the quantities N, $F_\nu$, Ra, or De (respectively pink, yellow, orange and blue). Datum from any of these child quantities are homogeneous, e.g. they have the same units and format) although the ensemble of datum for $Q_{set}$ is not.*

Nevertheless, a slice into the value of $Q_{set}$ at row 2, would return the *structure of $Q_{set}$* with each child quantity of that structure evaluated at row = 2, e.g.

$$Q_{set}(I_{row} = 2) \ = \ \{ \ N(I_{row} = 2), F_\nu,(I_{row} = 2), Ra(I_{row} = 2), De(I_{row} = 2) \ \} \quad (9)$$
$$+ \text{ meta-data of } Q_{set}$$

where the structure contains not only a list of the child quantities of $Q_{set}$ evaluated at row 2, but all of its meta-data too (which is not evaluated in terms of our cut)1.

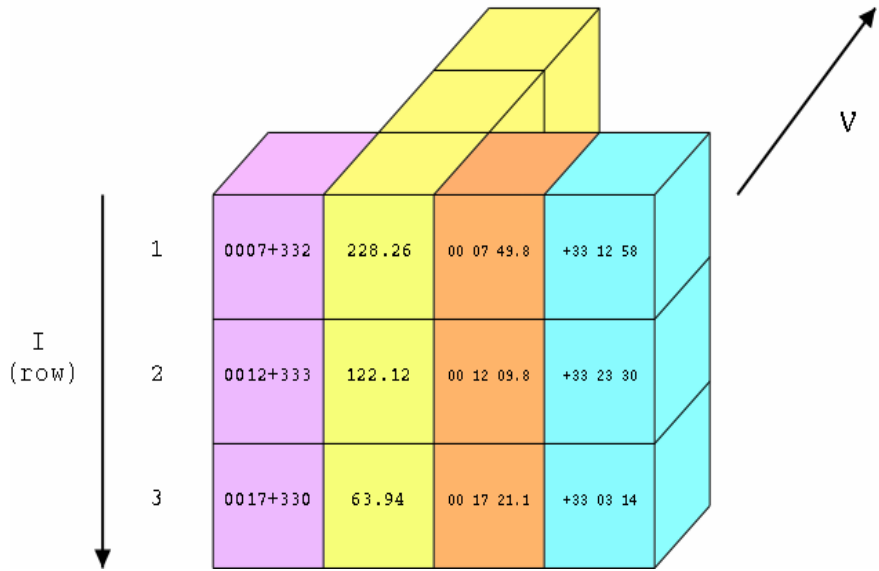

**Figure 10**. *Example showing how data cube of the quantity $Q_{set}$ may be non-symmetrically populated. Different values of ν only effect which datum is returned by the child quantity $F_ν$ (yellow cells).*

But we are not limited to just this simple layout of data in figure 9. Consider now that one of the quantities in the set $Q_{set}$ is itself dependent on another quantity. For example, the quantity "$F_ν$"might be a function of ν. We see then that the data cube would be that shown in figure 10 and that $Q_{set}$ is a function of both the quantities ν and $I_{row}$. For the majority of the quantities in $Q_{set}$ the dependent quantity ν is irrelevant to the datum returned (since N, RA, De do not depend on ν). For example, a query into the value of N(ν=10, $I_{row}$ = 2) is equivalent to N($I_{row}$ = 2). If, for the sake of argument, N did depend on ν then figure 10 would represent locations of N which are null (undefined).

Nevertheless it is clear that $Q_{set}$ *may* take as an argument on any quantity that its children are dependent on. To be clear about which quantities $Q_{set}$ actually depends on we should formally state which of its child arguments are an argument of $Q_{set}$. In doing so, then it means that the declared argument of $Q_{set}$ *is the same argument for all of its children* essentially meaning that *all child quantities are aligned along the same common dimension*. In the example in figure 10, $I_{row}$ is a common argument of the child quantities N, $F_ν$, RA, and De and thus an argument of $Q_{set}$ while the quantity ν is not.

---

1This is one functional definition of how quantities used as the data and meta-data differ (a slice through the parent on one or more of its dependencies doesn't affect the resulting meta-data returned. We could, of course, choose to ask for the list of the meta-data of $Q_{set}$ and then evaluate those child quantities in terms of the dependencies that the child meta-data quantities have.

Thus a slice through the $Q_{set}$ shown in figure 10 will result in a $Q_{set}$ with child quantities N, RA and De (evaluated at $I_{row} = 2$) containing *a single datum*, but child quantity $F_{if\mathcal{S}}$,($I_{row} = 2$) will consist of a linear array (with dependency on v).

## 2.5 Algorithmic generation of values

There is no reason that we cannot leverage this model to hold more than data in a file or database. Of particular interest are datum generated from an algorithm. To achieve this, we must insure that the generated data have consistent units, data format and accuracy for each generated d. For example the algorithm A, a function of the index i, creates values:

$$A(i) \mid i = 1.. \, n \quad = \{ \, d_1 \, .. \, d_n \, \} \qquad (10)$$

where n is the number of datum that the parent quantity holds. It is for the VO community to develop suitable recognized algorithms. For the purpose of illustration of this concept, we diagram how a "polynomial"might be created within the quantity framework (see figures 11 and A3).

## 3  Practical application

### 3.1 Development of domain-based ontologies

In practical application our data model must be realizable in software, thus we have described our model using the OWL [21] ontology language (figure 11). We choose OWL because it may be used to derive both the XML schema [17] and UML diagrams ([14]; see appendix 2 for UML diagrams of the quantity data model) which will serve as the basis for generating both code and XML instance documents. Furthermore, OWL, and ontologies in general, can be used to formally define the relationships between classes and other classes as well as their instances. Of course, placed on this footing we may now talk of quantities as classes in the object-oriented sense, which may be "extended" to create concepts that will be queried for and traded between participants in the VO. As a starting point, some important concepts that the community might choose to extend the quantities into in the VO domain include the space-time schema of Rots ([16]) and the UCD descriptors ([9]).

The development of ontologies that show the relationship between all of the concepts in the VO will be an important task and we believe that the quantity-based data model can form the basis for all the other VO ontologies. We need not, nor is it desirable, create a single all-encompassing ontology. It is better to have a spare, shared group ontology, working in tandem with richer, but localized ontologies. Mapping between localized ontologies can serve to bridge differences in defined concepts.

**Figure 11.** *Quantity data model ontology (in OWL)*

Each ontology should belong to its own namespace or domain as doing so has some benefits for the development of VO software. Consider that a concept might be created by a local VO participant "catalogID". Attaching the local namespace prevents collisions or confusion if another VO participant creates the same named concept. Having a "global"domain from which concepts may inherit allows for the possibility to still use the concept, at its globally defined level, even if the local meanings are not understood. Hopefully, as common concepts are discovered in local ontologies, they will be "promoted "up to the more global ontology. Having such a domain-oriented system will then allow freedom to create new concepts at the local level without breaking the functionality of the VO as a whole.

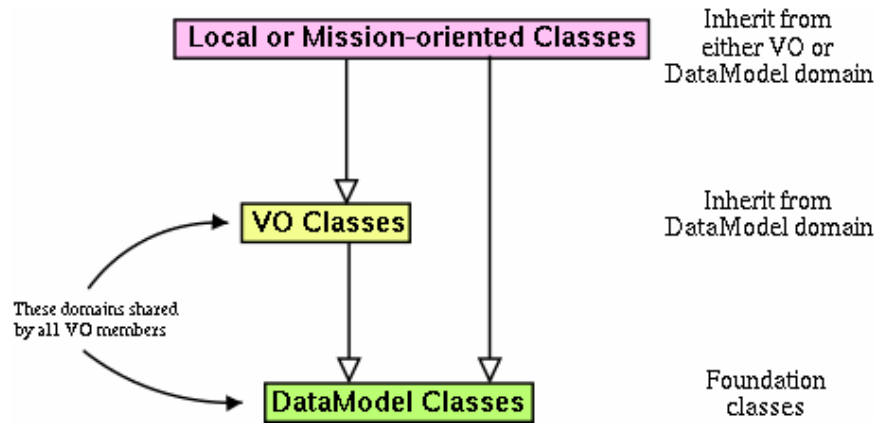

**Figure 13.** *Domains (namespaces) needed by model. The classes in the DataModel domain (Green) form the foundation of the model and are inherited by classes in both VO (Yellow) and Local (Pink) domains. The lines between colored boxes indicate the inheritance relationships, with hollow arrows pointing to the parent domain.*

We believe that something like a domain framework containing 3 levels of domains which are built one on top of the other (figure 13) is needed. At the base of the system is the "DataModel"domain which describes only *scientific* data and provides the framework for organizing this information. The quantity data model we have so far described lies within this domain. On ``top'' of this domain, and inheriting all of its concepts lies the ``VO'' domain which describes *astronomical* and VO-oriented technical concepts (such as meta-data to enable Grid technologies [13]). Finally, a number of ``Local'' domains will overlie the VO domain, and are present to enable the creation of machine-understandable standards that may occur amongst various sub-communities of the VO, but not the whole. Objects within the "VO"and "Local"domains may inherit from the DataModel::Quantity class.

### 3.1 Development of XML representation

Having an XML representation is important because it is the likely language that the data model will be used in for practical application. In XML, classes may be represented via XML schema [17] whereas any XML document represents actual instances of concepts.

Within the VO then it is easy to foresee that archives will want to develop shared XML schema/ontologies with which classes are defined for interchange and local XML documents, which use these schema/ontologies. A language similar to XQuery [19] might be utilized to interrogate these XML documents (catalogs of holdings) in terms of the concepts created by the quantity data model. While it is beyond the scope of this paper to really discuss how the search might be done, we refer to the work of [20,23] to give a useful survey of current thinking in terms of the present technologies.

After small consideration of the meta-data needs of XML (placement of id/idref attributes to allow for that 'pointer' mechanism in XML) we may use the UML software to relatively easily generate both code and an XML schema (see Appendix 1 for resources). We have mapped our model into an XML 'instance' using the data from table 1 in [18] (see figure 14; note that for reasons of space we had to truncate our example. A link to a fuller example may found in Appendix 1). To indicate how the XML mapping might work, we have colorized the locations of concepts in figure 14 which inherit from quantities using colored boxes based on the domain of the concept.

## 4  Discussion

With the model we have presented in this paper, we have managed to meet all of our set requirements and design goals. The model is sparse, with few classes, yet powerful enough to encompass many types of data that will exist in the VO. Furthermore, we have managed to cast this model in terms of an XML representation, which is important for transport of information across the Internet as well as its used in designing catalogs for searching holdings at data repositories.

Nevertheless, there are a number of critical issues which remain unsolved and open. First amongst them is the issue of data IO. Our model is designed with the intention that it be able not only to hold data, but to wrap existing files and data within databases yet we have not described how this might be achieved. This goal is important to the future of any data model as it is undesirable to have to transform all the data we want to make available in the VO from it present format. A crucial test of any data model will be whether or not an IO package can be developed to facilitate this type of functionality.

But more than this, it will be critical to show that any data model may have some binary representation for its data. XML, which is an ASCII language, will not do brilliantly at transporting large volumes of data. A prescription for holding or wrapping these large data volumes (preferably held in a binary representation) will need to be developed but we feel this is quite achievable, as it has already been tackled by a number of groups ([3] and [22] amongst others). We believe that these objectives may be reached by design of a complementary, but separate, IO package with minimal expansion of the meta-data contained in the data model. We hope to detail how this might be done in a future work.

```xml
<?xml version ="1.0"?>
<!DOCTYPE ADC:holding SYSTEM "dataset.xsd"[
  <!ENTITY J_ApJ_464_79_table1_data"SYSTEM "table1.fits"NDATA fits>
]>
<!-- ADC:holding inherits from VO:Collection, just adds a local id -->
<ADC:holding name="ApJ_464_79_table1"description=""adcId="ApJ_464_79"
            targetNamespace="http://www.ivoa.net/DataModel.xsd"
            xmlns:VO= "http://www.ivoa.net/VO.xsd"
            xmlns:ADC="http://www.adc.gsfc.nasa.gov/dataset.xsd"
>
  <VO:Origin><values><data>ADC</data></values></VO:Origin>
  <VO:Date><units><VO:date/></untis><values><data>2000-07
                        31T17:41:31</data></values></VO:Date>
  <members>
  <VO:Table name="table1: Canada-France Redshift Galaxies at z>0.5"
            description="ApJ/464/79: CFRS XI: High-redshift
            field galaxies morphology (Schade+ 1996)"
  >
    <argument>
       <index id="rowId"size="143"/>
    </argument>
    <members>

      <VO:survey name="CFRS"id="field1"description="Canada-France
                                    Redshift Survey designation">
          <argument><index idRef="rowId"/></argument>
          <unitless/>
          <values href="ApJ_464_79_table1_data"startByte="0"
                endByte="150">
            <floatDataFormat width="7"precision="4"/>
          </values>
      </VO:survey>
      …
      <VO:redshift  name="z"id="field2"description="Redshift">
          <argument><index idRef="rowId"/></argument>
          <unitless/>
          <values>
            <floatDataFormat width="6"precision="3"/>
            <data href="ApJ_464_79_table1_data"startByte="151"
                endByte="300"/>
          </values>
      </VO:redshift>
      …
      <quantity name="B/T"fieldId="field7"description="Colour versus
                bulge fraction">
          <argument><index idRef="rowId"/></argument>
          <unitless/>
          <values href="ApJ_464_79_table1_data"startByte="451"
                endByte="600">
            <floatDataFormat width="5"precision="2"/>
          </values>
      </quantity>
      …
    </members>
    <note>Created by Brian by hand. Many errors could exist!</note>
  </VO:Table>
  </members>
</ADC:holding>
```

***Figure 14.*** *XML example of quantity data model description of FITS data (taken from table 1 in [18]*

Another critical and unsolved issue is the fact that our model allows multiple ways to hold the same information. Imagine that a "sky coordinates"concept is created, which aggregates the "RA"and "De"concepts. Is it better to always use this object or perhaps to define a 'sky vector' with components RA and De. Clearly, one may ask the question if this functionality is a "bug or feature". Whatever the answer, we believe that at the least, a "good practices"document will have to be developed in order to fill the gaps in design that OWL, or other ontologies currently don't control well.

Lastly, there are a number of possible, common meta-data terms which we have neglected or insufficiently detailed in this work. Primary in this regard are our sparse description of the nature of the units and some of the listed objectives for data model meta-data as outlined by McDowell *etal*. ([2]). Some notable missing/unconsidered issues include how to describe the quality/fidelity, provenance, some important encoding schemes of data. Other issues include the support of some of the more unusual types of astronomical data including how to describe ranges of numbers, as well as censored data (e.g. data with upper or lower-limits). We hope to address many of these items this in another work.

## 5 Summary

We have produced a rudimentary data model based on theoretical considerations and mapped them into an ontological space. In addition, we have given thought to how this data model may be utilized in terms of software, and how the community would approach designing concepts for search and interchange.

### Acknowledgment

Support for this work was provided in part by NSF through Cooperative Agreement AST0122449 to the Johns Hopkins University.

# References

[1] McDowell, J., etal, 2002, ``Data Models for the VO: overview'', http://bill.cacr.caltech.edu/cfdocs/usvo-pubs/files/vodm003.ps

[2] McDowell, J., etal, 2002, ``Data Models for the VO: PartII: Metadata objects for the VO'', http://bill.cacr.caltech.edu/cfdocs/usvo-pubs/files/vodm003.ps

[3] XDF homepage: http://xml.gsfc.nasa.gov/XDF

[4] Plante, R., 2002, ``A Scalable Metadata Framework for the Virtual Observatory'', http://bill.cacr.caltech.edu/cfdocs/usvo-pubs/files/fw-draft2.pdf

[5] Plante, R., 2003, NVO Cambridge workshop presentation.

[6] Williams, R., etal., 2002, ``VOTable: A Proposed XML Format for Astronomical Tables'', http://bill.cacr.caltech.edu/cfdocs/usvo-pubs/files/VOTable-1-0.pdf

[7] VOTABLE homepage: http://www.vo-us.org/VOTable

[8] VO mailing list archive: http://archives.us-vo.org/dm

[9] UCD link: http://vizier.u-strasbg.fr/doc/UCD.htx

[10] Tody, D., etal., 2002, ``Simple Image Access Prototype Specification'',

http://bill.cacr.caltech.edu/cfdocs/usvo-pubs/files/ACF8DE.pdf
*[11]* FITS homepage: http://fits.gsfc.nasa.gov
*[12]  Astron. Astrophys. Suppl. Ser., Vol. 70, p.77, 1987*
*[13]* GRID homepage http://www.globus.org
*[14]* ``Unified Modeling Language'' resource page at: http://www.omg.org/uml
*[15]* XML resource page: http://www.w3c.org/XML
*[16]  Rots, A., 2003, ``Space-Time Coordinate Specification for VO Metadata'',*
 http://hea-www.harvard.edu/~arots/nvometa/SpaceTime.html
*[17]* XML schema  page: http://www.w3c.org/XML
*[18]  Schade etal., 1996, ApJ, 464, 79*
*[19] XQuery:  http://www.w3.org/TR/xquery*
*[20] Hoschek, Wolfgang, 2002, Ph.D. Thesis, University of Linz*
*[21] OWL : http://www.w3.org/TR/owl-features*
*[22] BinX : http://www.edikt.org/binx*
*[23] Dowler, P. etal., 2003, "VOQL", proceedings of ADASS XIII conference,
Strasborg, France.*

## Appendices

### Appendix I. Software resources

XML schema, an XML example that wraps a FITS file and simple Java code  may be found at the following URL:

   http://nvo.gsfc.nasa.gov/QuantityDataModel

### Appendix II. UML diagrams

In this document we will primarily present OWL, but here we include some UML diagrams for those of you who like them better. Minor discrepancies between these figures and the OWL exist.

*ABOUT figure notation and how these map to XML classes:*

In general, UML classes, which appear as boxes in the figures, generally map directly to nodes in the XML representation. Green is used to denote classes which belong to the ``DataModel'' domain, yellow is used to denote classes which belong to the ``VO'' domain. We will adopt the XML namespace to represent the class namespace, this means that  while in XML snippets the namespace will appear before a single colon, and in the UML diagrams  before a double colon the two are actually equivalent.

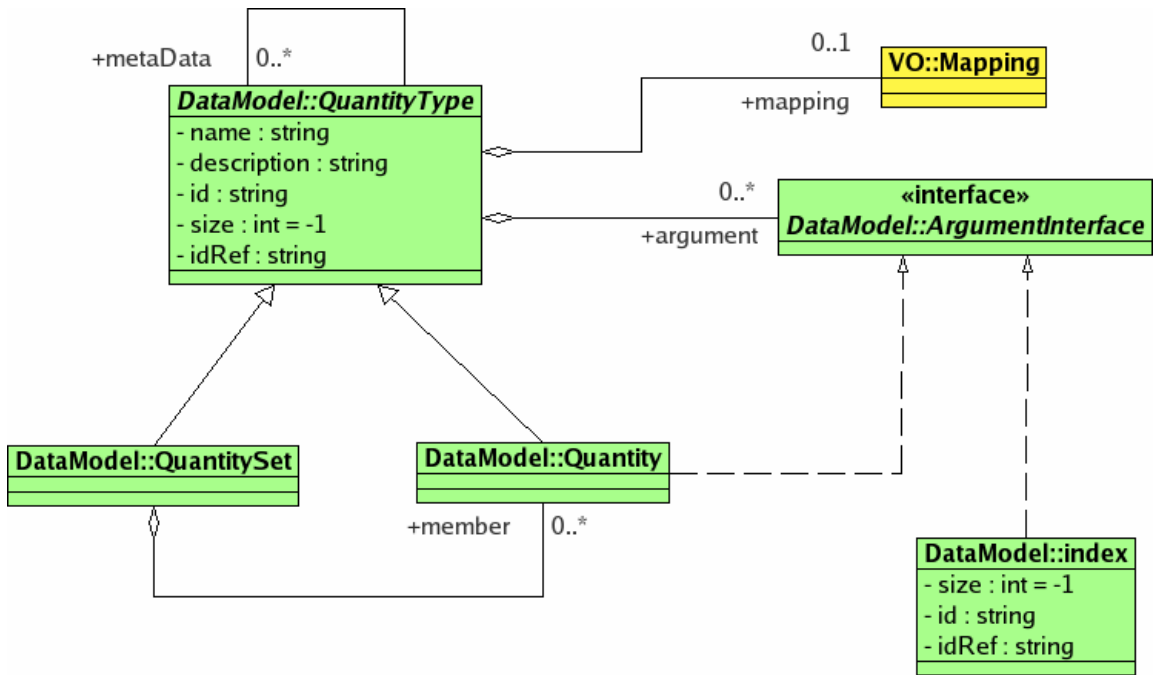

**Figure A1.** *The basic definition of what a quantity is, e.g. "QuantityType". We see both "Quantity"and "QuantitySet"are types of quantity. Index class exists to allow simple connection of a quantity to a dimension (e.g. the i,j,k,.. in the text). QuantitySet takes quantity as a member.*

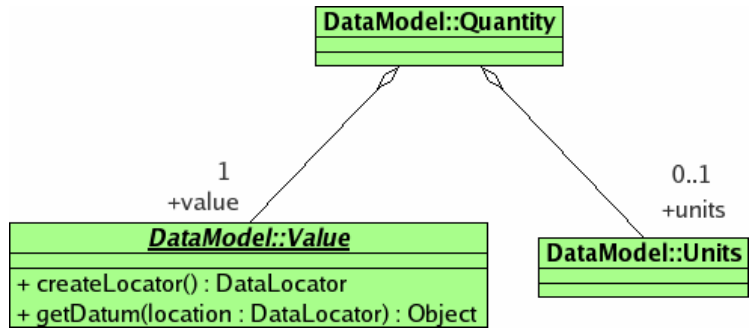

**Figure A2.** *The relationship of the quantity to its value and units.*

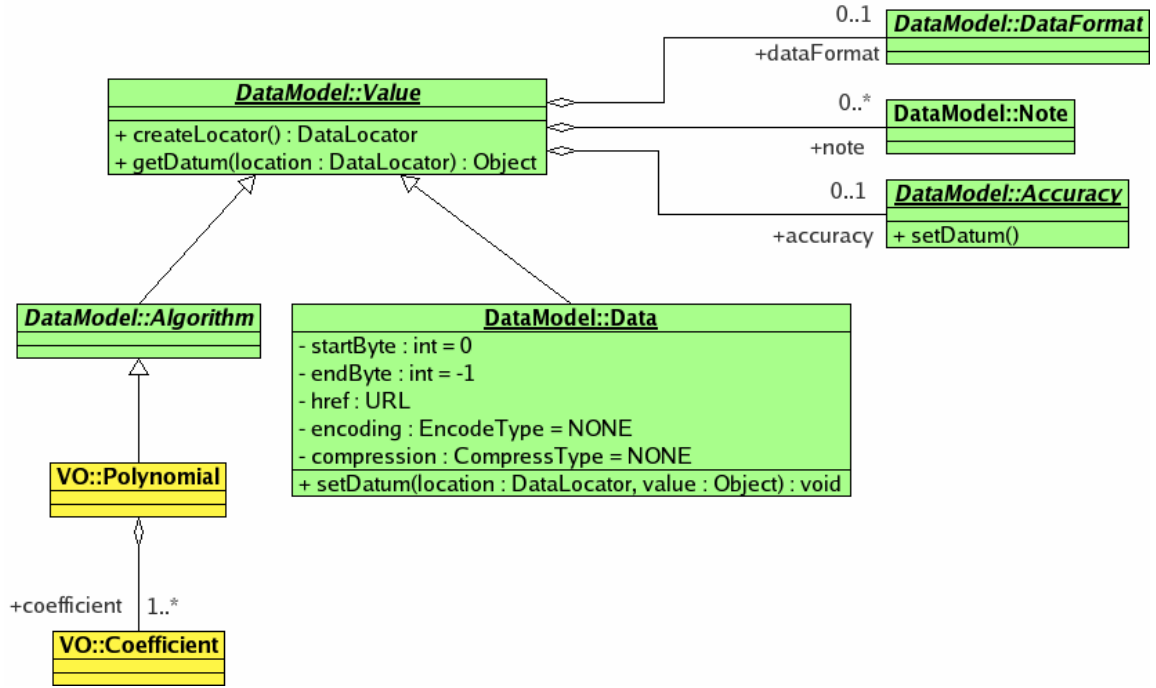

**Figure A3**. *Various sub-classes of data container.*

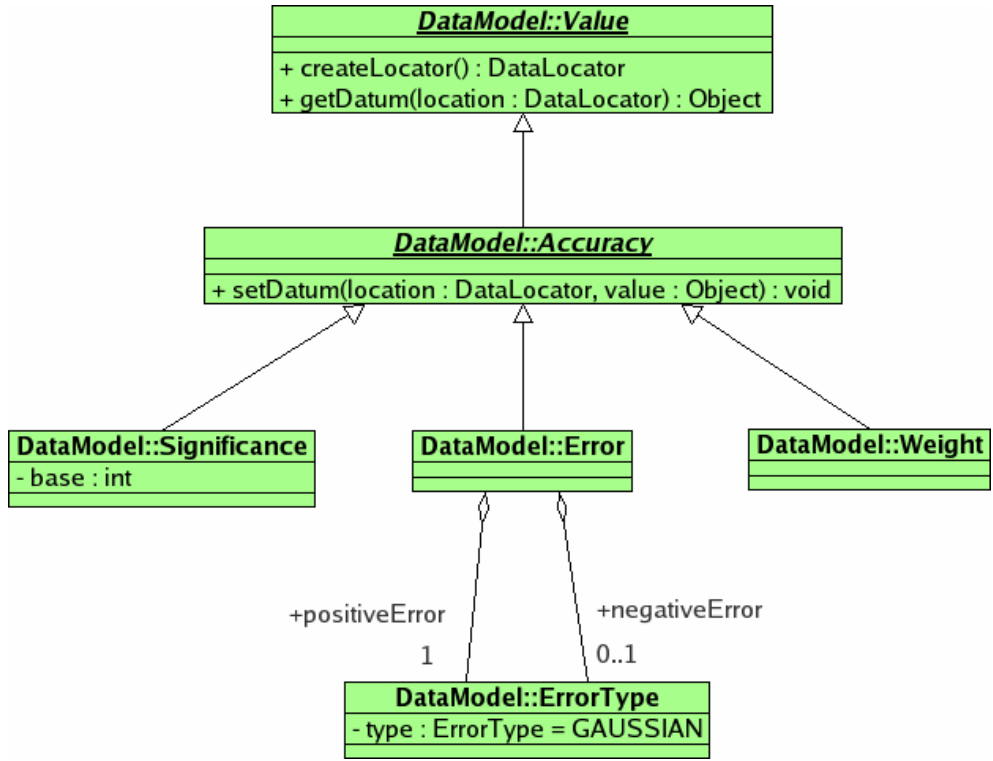

**Figure A4.** *Accuracy inheritance tree.*

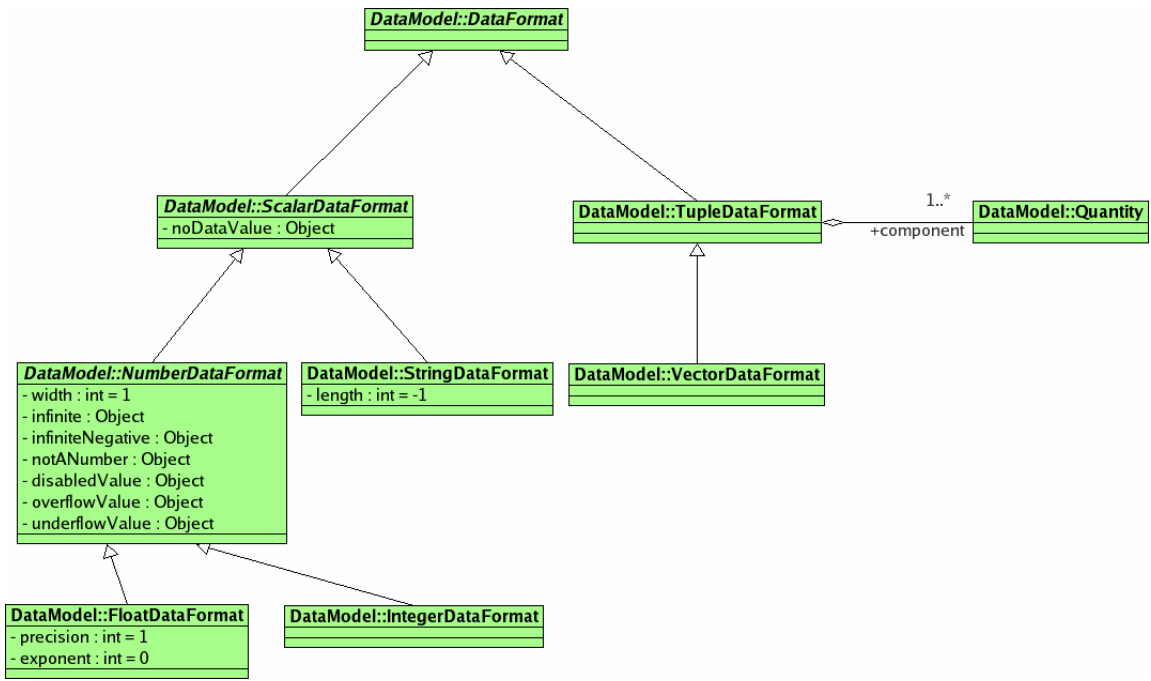

**Figure A5.** *Classes inheriting from data format.*